\documentclass[prc,twocolumn,showkeys,showpacs,superscriptaddress,]{revtex4}

\usepackage{dcolumn}
\usepackage{graphicx,color}

\usepackage{amsmath,amssymb}

\begin {document}
  \newcommand {\nc} {\newcommand}
  \nc {\beq} {\begin{eqnarray}}
  \nc {\eeq} {\nonumber \end{eqnarray}}
  \nc {\eeqn}[1] {\label {#1} \end{eqnarray}}
  \nc {\eol} {\nonumber \\}
  \nc {\eoln}[1] {\label {#1} \\}
  \nc {\ve} [1] {\mbox{\boldmath $#1$}}
  \nc {\mrm} [1] {\mathrm{#1}}
  \nc {\half} {\mbox{$\frac{1}{2}$}}
  \nc {\thal} {\mbox{$\frac{3}{2}$}}
  \nc {\fial} {\mbox{$\frac{5}{2}$}}
  \nc {\la} {\mbox{$\langle$}}
  \nc {\ra} {\mbox{$\rangle$}}
  \nc {\etal} {\emph{et al.\ }}
  \nc {\eq} [1] {(\ref{#1})}
  \nc {\Eq} [1] {Eq.~(\ref{#1})}
  \nc {\Ref} [1] {Ref.~\cite{#1}}
  \nc {\Refc} [2] {Refs.~\cite[#1]{#2}}
  \nc {\Sec} [1] {Sec.~\ref{#1}}
  \nc {\chap} [1] {Chapter~\ref{#1}}
  \nc {\anx} [1] {Appendix~\ref{#1}}
  \nc {\tbl} [1] {Table~\ref{#1}}
  \nc {\fig} [1] {Fig.~\ref{#1}}

  \nc {\bfig} {\begin{figure}}
  \nc {\efig} {\end{figure}}
  \nc {\ex} [1] {$^{#1}$}
  \nc {\Sch} {Schr\"odinger }
  \nc {\flim} [2] {\mathop{\longrightarrow}\limits_{{#1}\rightarrow{#2}}}
  \nc {\IR} [1] {\textcolor{red}{#1}}

  \nc {\gs} {g.s.}
  \nc {\es} {e.s.}

\title{Deducing spectroscopic factors from wave-function asymptotics}
\author{P.~Capel}
\email{pierre.capel@centraliens.net}
\affiliation{Physique Quantique, C.P. 165/82 and
Physique Nucl\'eaire Th\'eorique et Physique Math\'ematique, C.P. 229,
Universit\'e Libre de Bruxelles (ULB), B 1050 Brussels, Belgium}
\affiliation{National Superconducting Cyclotron Laboratory
and Department of Physics and Astronomy,
Michigan State University, East Lansing MI 48824, USA}
\author{P.~Danielewicz}
\email{danielewicz@nscl.msu.edu}
\author{F.~M.~Nunes}
\email{nunes@nscl.msu.edu}
\affiliation{National Superconducting Cyclotron Laboratory
and Department of Physics and Astronomy,
Michigan State University, East Lansing MI 48824, USA}
\date{\today}
\begin{abstract}
In a coupled-channel model, we explore the effects of the
coupling between configurations
on the radial behavior of the wave function and in particular on the
spectroscopic factor (SF) and the asymptotic normalisation coefficient (ANC).
We evaluate the extraction of a SF from the ratio of the
ANC of the coupled-channel model and that of a single-particle approximation
of the wave function.
We perform this study within a core$+n$ collective model, which
includes two states of the core that connect by a rotational coupling.
To get additional insights,
we also use a simplified model that takes a delta function for
the coupling potential.
Calculations are performed for \ex{11}Be.
Fair agreement is obtained between the SF inferred from the single-particle
approximation and the one obtained within the coupled-channel models.
Significant discrepancies are observed only for large coupling strength
and/or large admixture, i.e.\ small SF.
This suggests that reliable SFs can be deduced from
the wave-function asymptotics
when the structure is dominated by one configuration, i.e.\ for large SF.
\end{abstract}
\maketitle

\section{Introduction}
Historically, direct reactions have been an important
source of information on nuclear structure.
The development of radioactive-ion beams has rekindled
the interest in those reactions by providing a unique way
to study nuclei far from stability \cite{Ful10}.
Through direct reactions, it is indeed possible,
with an appropriate reaction model,
to extract quantitative information on the structure
of nuclei bound for only milliseconds~\cite{Rou10}.
For example, the orbital configuration of the valence nucleons
in a number of neutron-rich nuclei has been determined
through transfer \cite{TAB07},
knockout \cite{HT03} and breakup \cite{Nak99} reactions.

The abrupt nature of direct reactions leads to limited structural
changes in the involved nuclei.
Therefore, the cross sections for those reactions can
be expressed in terms of overlap functions, which, strictly,
involve overlap integrals of the full wave functions
for the initial and final states of the nuclei \cite{TN09}.
Information about overlap functions can be inferred
from direct reactions.
In particular, the square of its norm, called spectroscopic
factor~(SF), is supposed to be obtained
from a comparison between theory and experiment.
Some controversy exists about the use of SFs when discussing
nuclear structure \cite{FH02,FS10}.
Nevertheless, SFs are attractive in that they give an insight into the
complexity of the nuclear states involved in the reaction.
Moreover, they can be compared to results of electron
inelastic scattering and of shell-model calculations.
Alternatively, it has been suggested that, since direct reactions are mostly
peripheral, they probe only the tail of the overlap function \cite{MN05,CN07}.
Accordingly, only the asymptotic normalization coefficient (ANC)
of the overlap function should be extracted from data analysis.

Currently there is no fully microscopic basis for
calculating overlap functions for nuclei with mass $A > 10$,
let alone a reaction model that includes
these microscopic overlap functions.
Reaction data are usually analyzed within models where nucleons
are grouped into few clusters, and in which overlap functions
are substituted by wave functions obtained by solving a \Sch equation in which the interactions between the clusters
are simulated by local potentials.
As the primary application of this work is the study of nuclear structure
far from stability, we consider the case of two-cluster systems,
such as one-neutron halo nuclei, in which a neutron is loosely bound
to the core of the nucleus.
In that case, the overlap function is approximated
by a single-particle wave function, which describes the valence neutron
evolving in a mean-field potential representing
its interaction with the core.
The cross section obtained with this single-particle approximation
of the overlap function is
renormalized to match experimental data and,
from that renormalization, a SF or an ANC is deduced.
In this work, we examine the validity of these assumptions
by addressing the following questions:
First, how accurately can one substitute the exact overlap
function by a single-particle wave function?
Second, if only the ANC can be extracted from measurements, can
it be reliably related to the SF?

Within a fully microscopic description of the nuclei,
overlap functions satisfy a set of strongly coupled equations.
Formally, this set can always be reduced to one equation for the
single function of interest, evolving under the influence of an
effective single-particle Hamiltonian.
However, due to strong couplings between the different channels,
that Hamiltonian is going to be highly nonlocal,
in terms of its dependence on both position and energy.
For instance, it will describe absorption of the probability flux
from one channel into another channel.
This transfered probability flux may be absorbed from one side of the nucleus
and returned to the other side of the nucleus.
Obviously, this involves a significant spatial nonlocality,
up to the size of the nucleus.  The associated
energy dependence of the effective Hamiltonian may be tied to the time delay
in returning the flux to the original channel.
Needless to say that such type of effects cannot be properly accounted for
in terms of a simple local potential.
By contrast, coupling effects with exchange of probability flux
between channels may be adequately simulated at a semiquantitative level
in collective models.

In this work, we consider the rotational-coupling model developed by
Nunes \etal \cite{NTJ96,face}.
In that model, the core of the nucleus is described as a deformed rotor,
which can be in various excited states.
This leads to a set of coupled equations for the core-nucleon
wave functions similar to those obtained within microscopic models.
The rotational model is applied to \ex{11}Be, the archetypal one-neutron
halo nucleus. This nucleus is described as a neutron loosely bound
to a \ex{10}Be core which can be in both its $0^+$ ground state
and $2^+$ excited state \cite{NTJ96}.
Note that similar models
have been developed by other groups, see e.g.\ Refs.~\cite{EBS95,Vin95}.

In order to assess the validity of the single-particle approximation,
and to understand the influence of the couplings upon an overlap function,
its SF, and ANC, we carry out cross-comparisons within theory.
We analyze the more realistic rotational model
in terms of the single-particle approximation.
In particular we compute the ANCs within the rotational model
and the single-particle approximation and we deduce a SF from their ratio.
Confronting the deduced SF to that directly obtained
within the rotational model, we can infer the reliability of the
extraction of SFs from ANCs.
This method could prove very valuable if ANCs can be efficiently measured
from direct reactions, like transfer \cite{MN05} or breakup \cite{CN07}.
Note that besides the accuracy of the method,
the extraction of SFs from ANCs is subject to other uncertainties.
In particular, the geometry of the core-neutron potential is
not known \emph{a priori}.
In this work, we focus on the validity
of the method and disregard this latter uncertainty \cite{MN05,PNM07,MNM08}.
Throughout our analysis, we always use the same potential geometry
within both the rotational model and
the corresponding single-particle approximation.

Besides the aforementioned rotational model~\cite{NTJ96,face},
we also employ a simplified schematic model with
two overlap functions generated with square-well potentials
and coupled by a delta function.
This simplified model may be largely solved analytically,
which enables us to interpret qualitatively the results of the
rotational model.

An important issue not treated in the present work
is the effect of short-range correlations,
which can contribute to the lowering of SF.
Those correlations are often the focus of investigations
in many-body theory of nuclear matter~\cite{DB04}.
They generally affect the
overlap function within the nuclear volume,
just as do the couplings to collective states.
Hence, although not exactly included here,
short-range correlations will be qualitatively
simulated within the rotational model.

Direct reactions are a useful tool to study nuclear structure
far from stability. However, many uncertainties remain in the
analysis of measurements.
While significant advances have been made recently
on the side of reaction theory \cite{SNT06r,SNT06,Mat06,Manoli08,BCD09},
the models employed in the analysis
remain fairly schematic relative to the
full many-body problem.
On the structure side, the attention of microscopic models
has been primarily directed at the treatment of short-range
correlations. Traditionally, little attention was paid to the
asymptotic part of the wave functions~\cite{PW01,FVC08},
which is essential in the analysis of direct reactions.
More recently, there have been
works focusing on the asymptotic
behavior~\cite{QN09,Tim10,BN10}.
However, the task of establishing this
behavior may become very complicated for loosely-bound nuclei,
and even more if multiple-channel configurations are involved.
For these and other reasons, the single-particle model is likely to remain
a work horse in data analysis in the near future.
Therefore, assessing uncertainties associated with that model is important.

In the next section we provide the necessary theoretical background
for this study. We first introduce the notion of overlap functions and
their couplings. We then present the various models used in this work:
the single-particle model, the rotational model, and the
schematic delta-coupling model.
In \Sec{sec:results}, we provide quantitative results.
First, we study the realistic conditions for the case of $^{11}$Be.
Second, we explore the limits of the parameter space in the effort
to draw general conclusions.
Finally, we investigate these results within the more
schematic delta-coupling model.
Our work is summarized in \Sec{sec:summary}.

\section{Theoretical considerations}\label{theory}

\subsection{Overlap function, SF, and ANC}
Within direct-reaction theory, the cross sections for transfer, break-up,
or knockout are usually expressed in terms of overlap functions~\cite{TN09}.
The reaction is indeed expected to reveal the dominant
configuration within the projectile structure, such as $A \rightarrow B + n$.
Here and in the following, we consider a nucleus $A$
that exhibits a strong two-cluster structure:
a valence neutron $n$ bound to a core $B$.  To simplify matters, the symbols $A$ and $B$ will also stand for the mass numbers, i.e.\ $B=A-1$.

As an introductory step, we derive the equations satisfied by
the overlap wave function within a general many-body formalism.
We will specifically refrain from discussing the issues of spin,
some antisymmetrization effects and center-of-mass corrections.
With the exception of antisymmetrization, which cannot be
considered exactly in a collective model,
these will be accounted for in the practical calculations.

Formally, the microscopic Hamiltonian describing the
motion of the nucleons of nucleus $A$ reads:
\beq
H_A= \sum_{i=1}^{A} t_i
+ \sum_{j>i=1}^{A} v_{ij} \, ,
\eeqn{ea1}
where $t_i$ is the kinetic-energy operator for nucleon $i$
and $v_{ij}$ describes the interaction between nucleons $i$ and $j$.
For simplicity, we do not mention three-body interactions.
The states $\nu_A$ of $A$ are the eigenstates of $H_A$,
\beq
E_{\nu_A}^A \, \Psi_{\nu_A}(\ve{r}_1,\ldots,\ve{r}_{A})=
H_A \, \Psi_{\nu_A}(\ve{r}_1,\ldots,\ve{r}_{A}) \, ,
\eeqn{ea2}
where $\{\ve{r}_i\}$ are the coordinates of the nucleons of $A$.
Similarly, an $(A-1)$-body Hamiltonian, with eigenstates $\Phi_{\nu_B}$
of energy $E_{\nu_B}^B$, can be defined for the core $B$.
As mentioned earlier, the states of $A$ may present a strong $B$-$n$
cluster structure.
In that case, it is worthwhile to describe the relative motion between the
core in its state $\nu_B$ and the valence neutron
in terms of the overlap function $\psi_{\nu_A \, \nu_B}$,
which is nothing but the projection of $\Psi_{\nu_A}$
onto the wave function describing the core state $\Phi_{\nu_B}$,
\beq
\psi_{\nu_A \, \nu_B} (\ve{r})   = & \int
d\ve{r}_1 \cdots d\ve{r}_{A-1} \,
\Phi_{\nu_B}^*(\ve{r}_1,\ldots, \ve{r}_{A-1}) \nonumber \\
& \times
\Psi_{\nu_A}(\ve{r}_1,\ldots, \ve{r}_{A-1}, \ve{r}) \,,
\eeqn{ea3}
where $\ve{r}$ is the coordinate of the valence neutron relative to the core.

The spectroscopic factor is defined as the square of the norm
of the overlap function:
\beq
S_{\nu_A \, \nu_B} = \int d\ve{r} \, |\psi_{\nu_A \, \nu_B} (\ve{r}) |^2 .
\eeqn{ea4}
This SF represents the probability that, within the state $\nu_A$ of~$A$,
the neutron may be
found in a combination with the core $B$ in state $\nu_B$.
The SFs add up to unity over all the states of $B$,
including those in the continuum,
\beq
\sum_{\nu_B}S_{\nu_A \, \nu_B} = 1,
\eeqn{ea5}
since that sum represents the square of the norm of $\Psi_{\nu_A}$.

To obtain the formal equations satisfied by the overlap function,
we project both sides of \Eq{ea2} onto $\Phi_{\nu_B}$.
Taking account of \Eq{ea3}, we obtain
\begin{align}
E_{\nu_A}^A \, & \psi_{\nu_A \, \nu_B}  =
\langle \Phi_{\nu_B} | H_A | \Psi_{\nu_A}\rangle  \notag \\
 &=  (E_{\nu_B}^B + t_{\ve{r}}) \, \psi_{\nu_A \, \nu_B} + \sum_{\mu_B} V_{\nu_B \, \mu_B} \, \psi_{\nu_A \, \mu_B},
\label{ea6}
\end{align}
where $t_{\ve{r}}$ is the $B$-$n$ kinetic-energy operator
and
\beq
V_{\nu_B \, \mu_B} = \langle \Phi_{\nu_B} | \sum_{i=1}^{B} v_{in} |
\Phi_{\mu_B} \rangle
\eeqn{ea7}
are potential terms that couple
different possible configurations $\nu_A \, \nu_B$ within $\Psi_{\nu_A}$.
Let us now define the energy of the state $\nu_A$ relative to
the $B$-$n$ threshold (i.e.\ the negative of the separation energy) by
\beq
\varepsilon_{\nu_A}^{Bn} = E_{\nu_A}^A -E_0^B \, ,
\eeqn{ea7a}
and the core excitation energy by
\beq
\epsilon_{\nu_B}=E_{\nu_B}^B-E_0^B \,.
\eeqn{ea7b}
In Eqs.~\eq{ea7a} and \eq{ea7b}, $E_0^B$ is the ground-state energy of $B$,
hence $\varepsilon_{\nu_A}^{Bn} < 0$ for particle stable $A$
and $\epsilon_{\nu_B} \ge 0$.
The set of equations \eqref{ea6} can now be rewritten into the form
\begin{align}
(\varepsilon_{\nu_A}^{Bn} - \epsilon_{\nu_B})  \, \psi_{\nu_A \,   \nu_B} & =
 (t_{\ve{r}}+ V_{\nu_B \, \nu_B}) \, \psi_{\nu_A \, \nu_B} \notag\\
& \hspace*{1em} + \sum_{\mu_B\ne\nu_B} V_{\nu_B \, \mu_B} \, \psi_{\nu_A \, \mu_B} \, .
\label{ea8}
\end{align}
The picture behind this set of equations is that of a neutron
moving around $B$, which can be in its various states.
The possible quantum numbers for the $B$-$n$ relative motion in each
configuration $\nu_A \, \nu_B$ are determined by conservation
laws and by the quantum numbers of
the states $\nu_A$ and $\nu_B$.
In particular, the orbital angular momentum $l$ of the $B$-$n$ relative
motion is determined by parity conservation and angular-momentum couplings.

The diagonal potential element $V_{\nu_B \, \nu_B}$ in \Eq{ea8} represents the single-particle
potential for $n$ being in the field of $B$ in its state $\nu_B$.
The motion of $n$ is modified by the presence of other channels $\nu_A \, \mu_B$,
which couple to $\nu_A \, \nu_B$ through the non-diagonal potential elements $V_{\nu_B \, \mu_B}$.
In passing, we may note that, since the internucleon interactions $v_{ij}$ are at most weakly nonlocal, both the single-particle and
the coupling potentials \eq{ea7} will be at most weakly nonlocal in $\ve{r}$.  In addition, this weak nonlocality, combined with the short range of nuclear
interaction, will lead to a form factor for
the nuclear contribution to $V_{\nu_B \, \nu_B}$
that approximates the shape of the density of $B$.
Outside the volume of $B$, the nuclear contributions to the potentials
get suppressed and the wave equations simplify.
For a $B$-$n$ system, i.e. in absence of Coulomb,
the radial part of the overlap function of a configuration
with orbital angular momentum $l$ exhibits the asymptotic behavior
\beq
 \psi_{\nu_A \, \nu_B}(r) \flim{r}{\infty} C_{\nu_A \, \nu_B} \, i \, \kappa_{\nu_A \, \nu_B} \, h_l^{(1)}(i \, \kappa_{\nu_A \, \nu_B} r) \, ,
\eeqn{ea9}
where $\hbar \, \kappa_{\nu_A\nu_B} =
\sqrt{2\mu_{Bn}|\, \varepsilon_{\nu_A}^{Bn}-\epsilon_{\nu_B}|}$,
with $\mu_{Bn}$ the $B$-$n$ reduced mass, and
$h_l^{(1)}$ is a spherical Bessel function of the third kind \cite{AS70}.
The normalization constant $C_{\nu_A \, \nu_B}$ appearing in \Eq{ea9}
is the asymptotic normalization coefficient.
The function $h_l^{(1)}$ accounts for the effects of the centrifugal barrier
and behaves as $\exp(-\kappa_{\nu_A \, \nu_B} r)/\kappa_{\nu_A \, \nu_B} r$
for $\kappa_{\nu_A \, \nu_B} r\gg l$.
Hence, of all the different overlap functions,
the one corresponding to the ground state of $B$
(i.e.\ with $\epsilon_{\nu_B}=0$)
dominates the asymptotic behavior of $\psi_{\nu_A \, \nu_B}$, whereas
the overlap functions corresponding to high-lying states of the core
barely stick out of the volume of $B$.
In the presence of couplings, any channel 
that satisfies
the conservation laws could,
in principle, contribute to the spectroscopic factor.
However, in the case of loosely-bound systems,
much of the wave function is expected to be outside the volume of~$B$.
The dominance of the $B$ ground-state channel in the asymptotic region
suggests that this channel also dominates the contributions
to the SF.

Due to the couplings, the shape of the overlap function $\psi_{\nu_A \, \nu_B}$
differs from the solution of a single-particle \Sch equation
in which the nuclear $B$-$n$ interaction is simulated by a local potential
tied to the density of $B$.
This influence of the couplings upon the shape of the overlap function,
and in particular on the connection between SF and ANC,
is the focus of the present work.
As mentioned in the introduction, we do not solve the full many-body
problem \eq{ea2} but rather simulate the set of coupled equations \eq{ea8}
using a collective model
in which the core is described as a deformed rotor \cite{NTJ96}.
The results obtained within that rotational model will then be
compared to those of a mere single-particle model in order
to evaluate the sensitivity of the overlap function to the couplings.
For a qualitative understanding of our results, we also study a more
schematic model in which two channels are coupled through a delta interaction.

For completeness, we lay out in Appendix \ref{appendix_Ham}
the formal reduction of the set \eqref{ea8}
to a single-particle equation with a nonlocal effective Hamiltonian. 
The nonlocality is dominated by couplings to low-lying excitations of the core,
such as those investigated within the models here.
That reduction can be, in fact, applied
directly to the models we employ,
and it is bound to yield the same shape of overlap functions we obtain.

The different models considered here are presented in the following subsections.
Each of them is particularized to \ex{11}Be, the test case of our study.

\subsection{Single-particle approximation}\label{sp}

Within the single-particle (sp) approximation,
only one configuration is considered.
This comes down to neglecting the last terms in the r.h.s.\ of \Eq{ea8},
which therefore reduces to a mere one-body \Sch equation.
After the factorization of the overlap function into its radial
$\psi^{\rm sp}_{nlj}$ and spin-angular ${\cal Y}_{lj}$ parts,
the single-particle \Sch equation reads
\beq
(\varepsilon^{Bn}_{\nu_A}-\epsilon_{\nu_B})\psi^{\rm sp}_{nlj}(r)=
\left[ t_r^l+V_{\nu_Blj}(r)\right] \psi^{\rm sp}_{nlj}(r),
\eeqn{ea10}
where $n$ is the principal quantum number and
$j$ is the quantum number associated with the angular momentum
obtained from the coupling of the orbital angular momentum $l$
and the spin of the neutron.
In \Eq{ea10}, the radial part of the $B$-$n$ kinetic-energy operator reads
\beq
t_r^l&=& \langle{\cal Y}_{lj}|t_{\ve{r}}|{\cal Y}_{lj}\rangle\nonumber\\
&=&-\frac{\hbar^2}{2\mu_{Bn}}\left(\frac{d^2}{dr^2}+\frac{2}{r}\frac{d}{dr}
-\frac{l(l+1)}{r^2}\right).
\eeqn{ea10a}

As in the many-body case, the asymptotic behavior
of the single-particle wave function, normalized to unity, is
\beq
\psi^{\rm sp}_{nlj}(r) \flim{r}{\infty} b_{nlj} \,
i \kappa_{\nu_A\nu_B} \, h_l^{(1)}(i\kappa_{\nu_A\nu_B} r),
\eeqn{ea11}
where $b_{nlj}$ is the single-particle ANC.
From Eqs.~\eq{ea9} and \eq{ea11},
it is clear that the many-body overlap function is directly proportional
to the single-particle wave function for large $r$.
If a reaction probes only the ANC of the overlap function,
and the assumption is made that the couplings
to other states have little impact on the shape of the overlap function,
then the SF deduced on the basis of the single-particle model
amounts to
\beq
S^{\rm sp}_{nlj}=\left|\frac{C_{\nu_A\nu_B}}{b_{nlj}}\right|^2 \, .
\eeqn{ea12}

This single-particle approximation is tested here
in the case of \ex{11}Be. This typical one-neutron halo nucleus
has two bound states: a $1/2^+$ ground state and a $1/2^-$
excited state. If one assumes the \ex{10}Be core to be in its $0^+$
ground state, the corresponding single-particle orbitals are
$2s1/2$ and $1p1/2$, respectively.
The nucleus \ex{11}Be also exhibits a $5/2^+$ resonance, which
is usually reproduced within the $d5/2$ partial wave coupled to the
$0^+$ ground state of the \ex{10}Be core \cite{CGB04}.
If \ex{10}Be can be in its $2^+$ excited state,
then these orbitals couple to other configurations,
as described in the following subsections.

\subsection{The rotational model}\label{rotor}

We next consider a simple step beyond the single-particle approximation:
we use a collective model where the core is deformed and allowed
to excite \cite{NTJ96,face,EBS95,Vin95}.
The Hamiltonian of such a $B$-$n$ system reads
\beq
H^{\rm rot}_A=H_B+t_{\ve{r}}+V_{Bn}(\ve{r},\xi),
\eeqn{e1}
where $H_B$ is the internal Hamiltonian of the core,
and now the effective interaction between the core and the neutron depends
on the internal degrees of freedom of the core $\xi$.
In the following, each level of the core considered in the calculation
is identified by its spin $I$ and parity $\pi_B$. The corresponding
wave functions and energies are denoted by $\Phi_{I^{\pi_B}}$
and $\epsilon_{I^{\pi_B}}$, respectively
\beq
H_B\Phi_{I^{\pi_B}}(\xi)=\epsilon_{I^{\pi_B}} \Phi_{I^{\pi_B}}(\xi).
\eeqn{Hc}
The wave function of the $B$-$n$ system is expanded in terms of
these core eigenstates
\beq
\Psi^{\rm rot}_{J^\pi}=\sum_i\psi^{\rm rot}_i(r){\cal Y}_i(\hat{\ve{r}}) \Phi_i(\xi),
\eeqn{e2}
where the subscript $i$ represents all possible quantum numbers
that couple to the total angular momentum $J$ and parity $\pi$.
For clarity, we omit the coupling coefficients.
In \Eq{e2}, the part of the wave function describing the $B$-$n$ relative
motion, i.e.\ the equivalent of the overlap function \eq{ea3},
is split into its radial $\psi^{\rm rot}_{nlj}$
and spin-angular ${\cal Y}_{lj}$ parts.
Replacing Eqs.~\eq{e1} and \eq{e2} into the \Sch equation \eq{ea2},
one arrives at the coupled-channel equations \cite{NTJ96,face,EBS95}:
\beq
(\varepsilon^{Bn}_{J^\pi}-\epsilon_i)\psi^{\rm rot}_i(r)&=&
\left[t_r^l+V_{ii}(r)\right]\psi^{\rm rot}_i(r)\nonumber\\
&&+\sum_{j\ne i}V_{ij}(r)\psi^{\rm rot}_j(r),
\eeqn{e4}
where the potential matrix elements $V_{ij}$, responsible for the couplings
between various components of the wave function, are defined by
\beq
V_{ij}({r})=\langle\Phi_i(\xi){\cal Y}_i(\hat{\ve{r}})|V_{Bn}(\ve{r},\xi)|
{\cal Y}_j(\hat{\ve{r}})\Phi_j(\xi)\rangle.
\eeqn{e5}

The coupled equations \eq{e4} simulates
those satisfied by the microscopic overlap function \eq{ea8}.
In the rotor model, the core is described as
a deformed rotor \cite{NTJ96}. Within this model, the internal
coordinates of the core $\xi$ are the three Euler angles,
and the functions $\Phi_{I^{\pi_B}}$ are
the Wigner rotation matrices $D^{I}_{MK}$, with
$M$ and $K$ the projections of $I$
in the lab and core intrinsic frames, respectively.
For the application we have in mind, namely $^{11}$Be, we consider only the
first two states of the \ex{10}Be core: the $0^+$ ground state,
and the first $2^+$ excited state ($\epsilon_{2^+}=3.368$~MeV).
They are seen as the first two states of a rotational band of
intrinsic projection $K=0$.

Consistent with a deformed core, the interaction between the core and the neutron
is described by a deformed Woods-Saxon potential:
\beq
V_{WS}(\ve{r})=-V_l\left\{1+\exp\left[\frac{r-R(\theta,\phi)}{a}\right]
\right\}^{-1},
\eeqn{e6}
in which the depth $V_l$ may depend on the orbital angular momentum $l$.
The radius $R$ reads
\beq
R(\theta,\phi)=R_0[1+\beta Y_{20}(\theta,\phi)],
\eeqn{e7}
where $\beta$ characterizes the deformation of the core.
Since this deformation is responsible for the couplings between configurations,
$\beta$ is also called \emph{coupling strength} in the following.
On top of this deformed central term, we also include the usual
(undeformed) spin-orbit coupling term:
\beq
V_{SO}(\ve{r})=\ve{l}\cdot\ve{s}V_{SO}\frac{1}{r}
\frac{d}{dr}\left[1+\exp\left(\frac{r-R_{SO}}{a}\right)\right]^{-1}.
\eeqn{e8}
The radial components $\psi^{\rm rot}_{nlj}$
are found by solving \Eq{e4}, imposing bound-state boundary conditions
and unit normalization for $\Psi^{\rm rot}_{J^\pi}$ using the program {\sc face} \cite{face}.

Within this model, the wave functions of the $1/2^+$ ground state
and $1/2^-$ first excited state of \ex{11}Be comprise the following components
\begin{eqnarray}
\Psi^{\rm rot}_{1/2^+}&=&\psi^{\rm rot}_{2s1/2}({r}){\cal Y}_{s1/2}(\hat{\ve{r}})\Phi_{0^+}(\xi)\nonumber\\
&+&\psi^{\rm rot}_{1d5/2}({r}){\cal Y}_{d5/2}(\hat{\ve{r}})\Phi_{2^+}(\xi) \nonumber \\
&+&\psi^{\rm rot}_{1d3/2}({r}){\cal Y}_{d3/2}(\hat{\ve{r}})\Phi_{2^+}(\xi),\nonumber\\
\Psi^{\rm rot}_{1/2^-}&=&\psi^{\rm rot}_{1p1/2}({r}){\cal Y}_{p1/2}(\hat{\ve{r}})\Phi_{0^+}(\xi)\nonumber\\
&+&\psi^{\rm rot}_{1p3/2}({r}){\cal Y}_{p3/2}(\hat{\ve{r}})\Phi_{2^+}(\xi)\nonumber \\
&+&\psi^{\rm rot}_{1f5/2}({r}){\cal Y}_{f5/2}(\hat{\ve{r}})\Phi_{2^+}(\xi).\nonumber
\label{e9}
\end{eqnarray}
Asymptotically the radial behavior of $\psi^{\rm rot}_{nlj}$ is
identical to that of a single-particle wave function $\psi^{\rm sp}_{nlj}$
\beq
\psi^{\rm rot}_{nlj}(r) \flim{r}{\infty} C^{\rm rot}_{nlj} \,
i \kappa_{\nu_A\nu_B} \, h_l^{(1)}(i\kappa_{\nu_A\nu_B} r),
\eeqn{e10}
but with a different ANC $C^{\rm rot}_{nlj}$.
Indeed, whereas $\psi^{\rm sp}_{nlj}$ are normalized to unity,
$\psi^{\rm rot}_{nlj}$ have norms less than one due to their couplings
with the other components.
In addition, the radial dependence of $\psi^{\rm rot}_{nlj}$ may differ
from its single-particle approximation  because of these couplings.

Within this model, SFs can be calculated directly
\beq
S^{\rm rot}_{nlj}=\int_0^\infty |\psi^{\rm rot}_{nlj}|^2r^2dr.
\eeqn{e11}
Comparing these ``exact'' SFs with approximation~\eq{ea12}
provides a good test of the single-particle approximation used in most
of direct-reaction models.

\subsection{The delta function coupling model}
\label{delta-coupling}
To get a qualitative understanding of the results obtained in
the rotational model, we have developed a more schematic collective model
to describe the ground state of \ex{11}Be.
Square wells are used as single-particle potentials and a delta function
simulates the coupling term.
For simplicity, the spin of the neutron is neglected.
We consider two radial components $\psi^\delta_0$ and $\psi^\delta_2$
corresponding, respectively, to the valence neutron in the $s$ wave coupled to
the $^{10}$Be core in its $0^+$ ground state, and to the valence neutron
in the $d$ partial wave
coupled to the first excited state $2^+$ of \ex{10}Be.
In this schematic model, the set of coupled equations \eq{ea8} reduces to
\beq
\varepsilon^{Bn} \, \psi^\delta_0(r)&=&
[t_r^0 + V_{Bn}(r)] \, \psi^\delta_0(r)\nonumber\\
&&+V_\delta(r) \, \psi^\delta_2(r)  \, ,\nonumber\\
(\varepsilon^{Bn}-\epsilon_{2^+}) \, \psi^\delta_2(r)&=&
[t_r^2 + V_{Bn}(r)] \, \psi^\delta_2(r)\nonumber\\
&&+V_\delta(r) \, \psi^\delta_0(r) \, ,
\eeqn{cc-delta}
where $V_{Bn}$ is a square well of radius $R_0$ and depth $V_0$,
and $V_\delta(r)=-\beta \, V_{0} \, R_{0} \, \delta(r-R_0)$
with a coupling strength~$\beta$.

Outside the potential, at $r > R_0$, the neutron wave functions
exhibit the asymptotic form
\begin{eqnarray}
\psi^\delta_0 (r) =  C^\delta_0 \, i \kappa_0 \, h^{(1)}_{0}(i\kappa_0 r) \, ,
\nonumber \\
\psi^\delta_2 (r) = C^\delta_2 \, i \kappa_2 \, h^{(1)}_{2}(i\kappa_2 r) \, ,
\label{delta-asymptotic}
\end{eqnarray}
where we have introduced the corresponding ANCs $C^\delta_0$ and $C^\delta_2$.
Inside the potential well, at $r < R_0$, the wave functions are
\begin{eqnarray}
\psi^\delta_0 (r) =  A_0 \, j_{0}(k_0 r) \, ,
\nonumber \\
\psi^\delta_2 (r) = A_2 \, j_{2}(k_2 r) \, ,
\label{delta-inside}
\end{eqnarray}
where $A_0$ and $A_2$ are normalization constants,
$j_l$ are spherical Bessel functions of the first kind \cite{AS70},
$\hbar k_0 = \sqrt{2 \mu_{Bn} (\varepsilon^{Bn} + V_0)}$, and 
$\hbar k_2 = \sqrt{2 \mu_{Bn} (\varepsilon^{Bn} - \epsilon_{2^+} + V_0)}$.
At $r=R_0$, the overlap functions $\psi^\delta_0$ and $\psi^\delta_2$
are continuous, but their derivatives are not, due to the delta coupling.
This discontinuity is proportional to the value of
the wave function in the other channel.

The conditions at $r=R_0$ may be combined into one equation relating $\varepsilon^{Bn}$, $V_0$ and $\beta$:
\begin{align}
\Big\lbrace\big[ & \log{(r \,  \psi^\delta_0 )}\big]' \big|_{R_{0+}}  - \big[\log{(r \, \psi^\delta_0 )}\big]' \big|_{R_{0-}} \Big\rbrace \notag \\
& \hspace*{1.5em} \times 
\Big\lbrace\big[\log{(r \, \psi^\delta_2 )}\big]'  \big|_{R_{0+}}  - \big[\log{(r \, \psi^\delta_2 )}\big]' \big|_{R_{0-}} \Big\rbrace \notag \\ &  \hspace*{4em}  = \Big( \frac{2 \mu_{Bn} \, \beta \, R_0 V_0 }{\hbar^2} \Big)^2 \, .
\label{eq:condition}
\end{align}
Here, $R_{0-}$ and $R_{0+}$ stand for limits taken
from below and above $R_0$.
At the extreme of a vanishing coupling, $\beta =0$,
the equation \eqref{eq:condition} will be satisfied when either of
the l.h.s.\ factors vanishes.  Vanishing of either of those factors
is indeed the standard square-well condition for a bound state,
when no coupling exists.  

In the presence of a coupling, we find it easiest
to fix $V_0$, which sets
the value for the l.h.s.\ of \eqref{eq:condition},
and read off $\beta$ from the r.h.s.
Thereafter, we can determine the ratios of the constants
$A_l$ and $C_l$ and, finally, absolute values for those constants
from normalization of the wave function to unity.

Besides the fact that it can be solved analytically,
the delta-coupling model presents the advantage that
it enables one to easily interpret the
features of the set of coupled equations \eq{cc-delta}
in terms of flux of probability transfered from one channel to the other.
As the wave for a given  $l$ encounters the boundary of the potential
generated by  the core, angular momentum may be exchanged with the core
and flux of probability can be transferred between the channels.
Of course the net probability flux is conserved within the whole
wave function, but not inside a single channel:
When starting from a solution for $l=0$, without coupling,
and then enhancing the coupling, the picture is that of the
coupling acting as an antenna, or source term,
that radiates into the $l=2$ channel.
For a wave traveling onto the coupling,
the flux from the $s$ channel to the $d$ one is
\beq
\lefteqn{R_0^2 \frac{\hbar}{\mu} \Im\left\{\psi^{\delta *}_0
\left(\psi^{\delta\prime}_0\big|_{R_{0-}}-\psi^{\delta\prime}_0\big|_{R_{0+}}\right)\right\}}\nonumber \\
&& =R_0^2 \frac{2}{\hbar}\beta R_0 V_0 \Im\left\{\psi^{\delta *}_2\psi^\delta_0\right\}|_{R_{0}},
\eeqn{ez1}
which shows that the discontinuity in the derivative of the wave function
due to the delta coupling is related to the transferred flux of probability.
In the stationary states we are interested in,
the flux leaking out of a channel
is exactly balanced by the flux coming in.

The SFs are the contributions from the two components
to the square of the norm of the wave function,
\beq
S^\delta_l=\int_0^\infty |\psi^\delta_l|^2 r^2 dr  \, .
\eeqn{delta-sf}
Again, we can compare those SFs
coming directly from the solution of Eq.~(\ref{cc-delta}) with their
single-particle approximation~\eq{ea12}.
The change in the shape of the wave function,
compared to the single-particle approximation,
is due to the fact that the derivative is discontinuous at $r = R_0$.
From the set \eqref{cc-delta}, it follows that the magnitude
of the $l=2$ wave function is linear in the coupling $\beta$, for small $\beta$.
With this, the discontinuity in the derivative of $l=0$ wave function
is quadratic in $\beta$.  In consequence, any discrepancy
between the exact spectroscopic factor and that estimated
in the single-particle approximation should be quadratic in the coupling.
At the same time, deviation of the spectroscopic factor from unity, for $l=0$,
should be quadratic in the coupling as well.
In the end, the discrepancy in the spectroscopic factor is expected to be
linear in the deviation of the dominant spectroscopic factor from unity.  

\section{Results}
\label{sec:results}

\subsection{Realistic $^{11}$Be}\label{real11Be}

There are three well known states in $^{11}$Be to which we fit the
$B+n$ interaction: the $1/2^+$ ground state (\gs) with an energy
$\varepsilon^{Bn}_{1/2^+}=-504$ keV,
the $1/2^-$ first excited state (\es) with $\varepsilon^{Bn}_{1/2^-}=-184$ keV,
and the $5/2^+$ resonance located at $\varepsilon^{Bn}_{5/2^+}=1.274$~MeV in
the continuum.
Our starting point for the undeformed interaction
(i.e.\ the uncoupled case) is the potential developed
in \Ref{CGB04}. The radius and diffuseness of the Woods-Saxon form factor are
chosen equal to $R_0=2.585$~fm, and $a=0.6$~fm, respectively.
The spin-orbit depth is fixed at $V_{SO}=21.0$~MeVfm\ex{2},
and its radius at $R_{SO}=2.585$~fm.

We wish to explore the effect of the single-particle approximation
as a function of the coupling strength.
Thus, although the quadrupole deformation $\beta$ has a definite value
($\beta=0.67$),
which is related to the $B(E2;0^+\rightarrow 2^+)$ of \ex{10}Be \cite{NTJ96},
we use it as a parameter. For each $\beta$,
the depths of the potential in the $s$ wave ($V_s$)
and $d$ wave ($V_d$) are adjusted
to reproduce the energy levels of the two positive-parity states.
The depth is chosen the same in all the odd waves ($V_p$) and is adjusted
to the energy of the \es\ The radius $R_0$ is modified as a function
of $\beta$ to conserve the volume of the core.
The spin-orbit coupling term is kept constant.

\begin{figure}
\includegraphics[width=8cm]{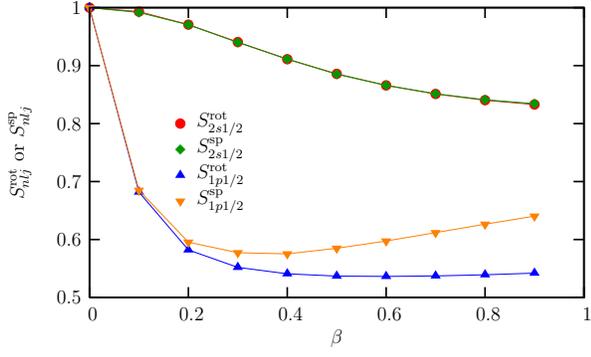}
\caption{(color online) Spectroscopic factors $S^{\rm rot}_{nlj}$ \eq{e11}
and single-particle approximation $S^{\rm sp}_{nlj}$ \eq{ea12}
as a function of deformation.
Note that the diamonds are superimposed on the circles.}
\label{f1}
\end{figure}

In Fig.\ref{f1} we show the spectroscopic factors $S^{\rm rot}_{nlj}$
obtained from
the solution of the coupled-channel equations as a function of deformation.
Results for the \gs\ (circles)
and the \es\ (up triangles) are both included.
The effect of deformation is very different
in these two states although they are both loosely bound, and
both correspond to a mixing of three configurations.
The $S^{\rm rot}_{2s1/2}$ in the \gs\ remains above $80$\%
even at large deformation.
The remaining strength is shared between both $d$ configurations.
On the contrary the $S^{\rm rot}_{1p1/2}$ in the \es\ drops rapidly
with deformation down to 50--60\%. The remaining part
of the strength is exclusively in the $p3/2$ configuration.
No strength is found in the $f5/2$ configuration.
These results are in agreement with those of Refs.~\cite{NTJ96,EBS95,Vin95}.
For both states, there is no configuration crossing:
only one configuration dominates for all deformations.
In both cases, the dependence of SFs on $\beta$ is strongly non-linear.
Interestingly, at small $\beta$, and when there is little admixture,
$1-S^{\rm rot}_{2s1/2}$ behaves roughly quadratically in $\beta$,
as discussed at the end of \Sec{delta-coupling}.

We compare these SFs with the values that would have been obtained under
the single-particle approximation \eq{ea12}.
For the \gs, this approximation (diamonds) works very well:
no matter how large the coupling strength,
$S^{\rm sp}_{2s1/2} \approx S^{\rm rot}_{2s1/2}$.
For the \es\ (down triangles),
only the general trend of $S^{\rm rot}_{1p1/2}$
is reproduced by $S^{\rm sp}_{1p1/2}$.
Although the single-particle approximation is valid at small deformation
significant discrepancies appears for $\beta>0.2$.

\begin{figure}
\includegraphics[width=8cm]{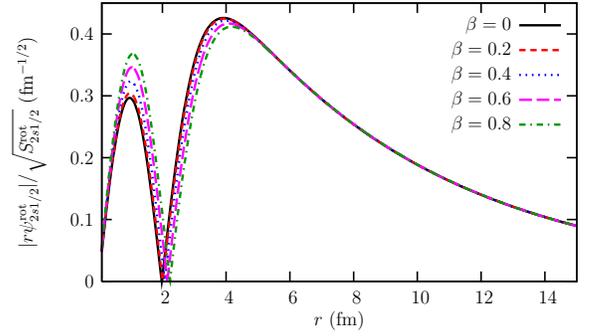}
\caption{(color online) Radial part of $2s1/2$ overlap functions
normalized to unity.}
\label{f2}
\end{figure}
\begin{figure}
\includegraphics[width=8cm]{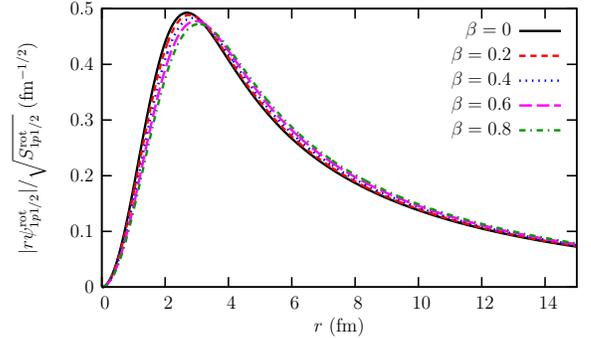}
\caption{(color online) Radial part of $1p1/2$ overlap functions
normalized to unity.}
\label{f3}
\end{figure}
These results can be further illustrated by plotting the $2s1/2$ and $1p1/2$
components of the wave function for various deformation parameters (Figs. \ref{f2}
and \ref{f3}).
These radial functions have been normalized to one to make the comparison
easier. We see that the couplings simply rearrange the interior
of the $s1/2$ contribution: increasing $\beta$ moves strength
from the second peak to the first. However the asymptotic part (beyond 5~fm)
is left totally unchanged, explaining the validity of
the single-particle approximation \eq{ea12}.
For this loosely-bound $1/2^+$ state, we find
that 50\% of the spectroscopic strength comes from radii smaller than
the interaction radius.
This illustrates
that even for loosely-bound systems, the internal part of the overlap function
significantly contributes to the SF.
The validity of the single-particle approximation
is therefore surprising.
Although couplings significantly affect the interior of the overlap function,
which contributes to a large part of the SF, the approximation \eq{ea12},
based solely on asymptotic characteristics of the wave function,
gives a precise estimate of the SF.

As opposed to the $1/2^+$ state, the couplings seem to affect the
wave function of the $1/2^-$ state all the way out to large distances
(see \fig{f3}).
In this case, strength is moved outwards when $\beta$ increases.
This explains the deviation between $S^{\rm rot}_{1p1/2}$ and its
single-particle approximation $S^{\rm sp}_{1p1/2}$ observed in \fig{f1}.

These results obtained for a realistic
description of \ex{11}Be are rather surprising.
First, it was expected that at some point,
e.g. for large deformation, the single-particle
approximation would break down.
However, as shown in \fig{f1}, the single-particle approximation
seems to be valid even at large $\beta$,
especially for the ground state.
Second, large differences are observed between the two states.
For the ground state, the $2s1/2$ configuration
dominates and the single-particle approximation is nearly exact.
On the contrary, the excited state presents a large
admixture of the $1p1/2$ and $1p3/2$ configurations,
and the single-particle approximation of its SF is less precise.

The difference in the validity of the single-particle approximation \eq{ea12}
can be understood in terms of
the different admixtures that are obtained in the \gs\ and the \es\
This suggests that the single-particle approximation is less
valid at large admixture, i.e.\ when the corresponding SF is small.
This agrees, at least qualitatively, with the arguments 
expounded in the description of the delta-coupling model.

One possible explanation for the difference in the configuration admixture
between both states is the presence
of a bound state in the $p3/2$ partial wave generated by
the core-neutron potential \cite{CGB04}.
The existence of such a state could attract probability flux
to that configuration and so enhance the admixture in the \es\ 
Accordingly, the low admixture observed in the \gs\ would be explained
by the absence of bound state in the $d$ waves.
To better understand these results, we perform similar
calculations extending the range of parameters to extreme values.

\subsection{Extreme couplings}\label{Xtreme}

\begin{figure}
\includegraphics[width=8cm]{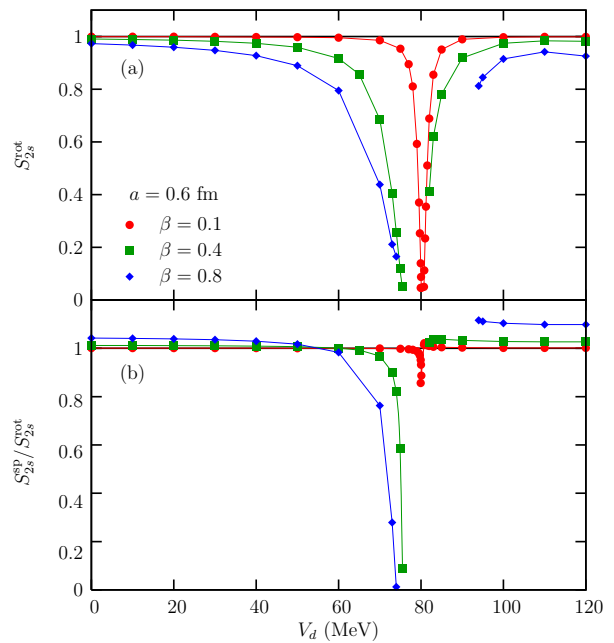}
\caption{(color online) (a) Spectroscopic factors as a function of $V_d$.
(b) Relative effect of the single-particle approximation as
a function of $V_d$. ($a=0.6$~fm)}
\label{f4}
\end{figure}

To study the difference in admixture observed
between the two bound states, we perform calculations for a hypothetical
\gs\ of \ex{11}Be with different values of $V_d$ to see
if and when the presence of a bound state in the $d$ wave
affects the admixture in that state.
To do so, we repeat the analysis performed in \Sec{real11Be} for the \gs,
but now, instead  of adjusting $V_d$ to the resonant state,
we use it as a free parameter
and adjust $V_s$ to reproduce the correct \gs\ energy.
For simplicity, the spin of the neutron is neglected.
Results are shown in \fig{f4}(a)
for deformation parameters $\beta=0.1$ (circles), 0.4 (squares),
and 0.8 (diamonds).
The most striking feature is
the pronounced drop of $S^{\rm rot}_{2s}$ at $V_d \approx 80$ MeV.
At this depth, the single-particle potential hosts a $d$ bound state
with an energy that corresponds approximately
to $\varepsilon^{Bn}_{1/2^+}-\epsilon_{2^+}$.
In that condition, we have a nearly degenerate system, where the
\gs\ can be simultaneously in both $2s$
and $1d$ configurations.
It is only the presence of this deep $d$ bound state that explains
the large admixture. Indeed at lower or larger $V_d$, the admixture
vanishes and $S^{\rm rot}_{2s}>0.8$.
This behavior holds for all coupling strengths,
though the region of large admixture increases with deformation $\beta$.
Note also the discontinuity of our results, especially at large $\beta$.
In the large-admixture range, parameters cannot be adjusted to
reproduce the binding energy of the system due to numerical limitations.

This result confirms the hypothesis formulated at the end
of \Sec{real11Be}. 
It shows that large admixture is to be expected near degeneracy,
i.e. when the coupled wave hosts a bound state close to the energy
$\varepsilon^{Bn}_{J^\pi}-\epsilon_{I^{\pi_B}}$.
However, if that bound state is located above or below that
energy, no significant admixture should be observed.
This explains the small admixture observed in the \gs\
of the realistic \ex{11}Be. In that case $V_d$ is adjusted to reproduce
the $5/2^+$ resonance. In the single-particle potential, all $d$
states are thus in the continuum, far from the energy at which there
would be degeneracy.
The admixture observed in the \es\ of the realistic \ex{11}Be
is also explained by this effect. The single-particle parameters indeed
lead fortuitously to a $1p3/2$ state bound by about $3$~MeV \cite{CGB04}, 
which corresponds approximately to $\varepsilon^{Bn}_{1/2^-}-\epsilon_{2^+}$,
i.e. the energy at which there is degeneracy.
Note that the large admixture observed in the \es\ of \ex{11}Be
is unphysical as the $p3/2$ orbital is Pauli blocked to the valence
neutron \cite{SBE93}.
Because of the large centrifugal barrier, the $1f5/2$ state is located
high in the continuum, and so does not couple to the $p$ configurations.

Besides aiding to understand the results
obtained in \Sec{real11Be},
these manipulations enable us to induce large admixture in
our two-cluster rotational model.
In this way we can study the validity of the
single-particle approximation in nuclei
with large fragmentation of strength
that usually require more advanced structure models.
In \fig{f4}(b), the ratio of the single-particle estimate
$S^{\rm sp}_{2s}$ \eq{ea12} to its
``exact'' value $S^{\rm rot}_{2s}$ \eq{e11},
is plotted as a function of $V_d$.
The approximation works perfectly at small coupling strength
and its validity decreases as $\beta$ increases.
However, no matter how large the coupling strength, the
single-particle approximation remains valid outside the large-admixture range
with less than 10\% error.
This confirms that the single-particle approximation breaks down
mostly at large admixture.

\begin{figure}
\includegraphics[width=8cm]{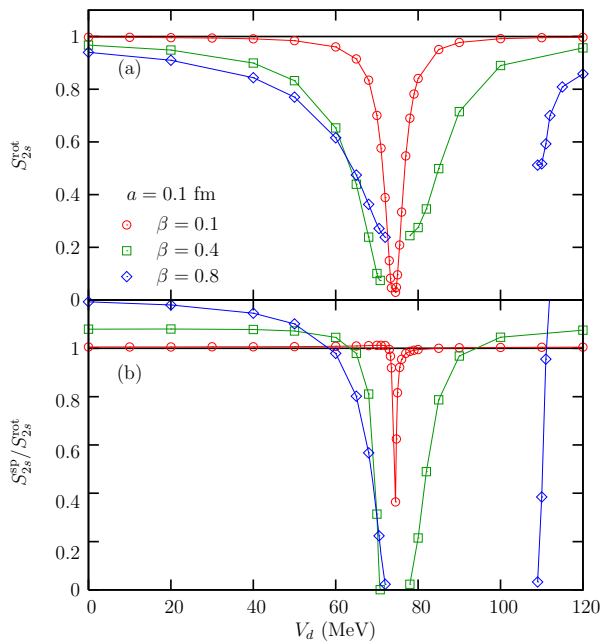}
\caption{(color online) (a) Spectroscopic factors as a function of $V_d$.
(b) Relative effect of the single-particle approximation as
a function of $V_d$. ($a=0.1$~fm)}
\label{f6}
\end{figure}

\fig{f6} shows the equivalent of \fig{f4}
but for a much smaller diffuseness $a=0.1$ fm.
We perform this set of calculation not only to
test the validity of our conclusions in extreme cases, but also to ease
the comparison between this rotational model and the delta-coupling
model, which corresponds to a nil diffuseness (see \Sec{deltanalysis}).
In this case, the large-admixture range is increased.
Therefore the breakdown of the single-particle approximation
happens sooner and to a larger extent.
This is due to the smaller diffuseness, which produces a more abrupt change
in the radial behavior of the wave function at the surface
and so enhances the couplings between the various configurations.

\begin{figure}
\includegraphics[width=8cm]{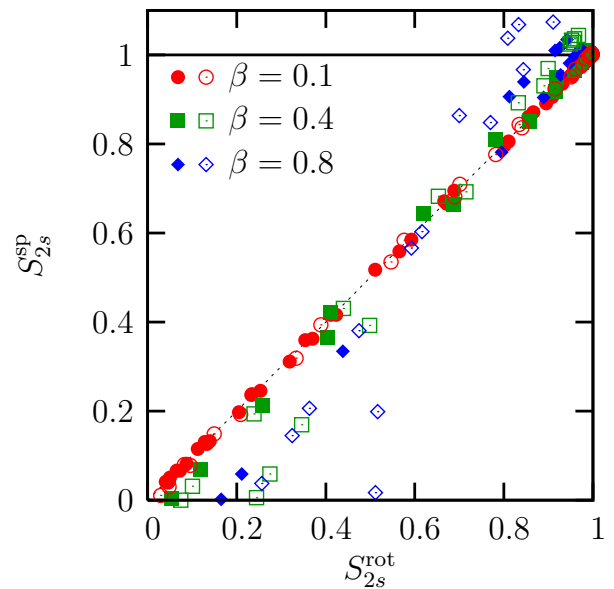}
\caption{(color online) Comparing the single-particle approximation
$S^{\rm sp}_{2s}$ to the ``exact'' $S^{\rm rot}_{2s}$.
Closed symbols correspond to diffuseness $a=0.6$~fm, and open ones to $a=0.1$~fm.}
\label{f7}
\end{figure}

To summarize this analysis, we plot in \fig{f7}
the single-particle approximation $S^{\rm sp}_{2s}$
against the ``exact'' $S^{\rm rot}_{2s}$
for all the cases explored in this section.
These correspond to the
deformations $\beta=0.1$, 0.4, and 0.8 (circles, squares,
and diamonds, respectively) and two diffusenesses $a=0.1$, and 0.6~fm
(open and solid symbols). The dashed line corresponds to
$S^{\rm sp}_{2s}=S^{\rm rot}_{2s}$ and the deviation from this line
estimates the error of the single-particle approximation.
The main information conveyed by \fig{f7} is
the general agreement between the ``exact'' SF and its
single-particle approximation.
Although there are some deviations from the $S^{\rm sp}_{2s}=S^{\rm rot}_{2s}$
line, a large $S^{\rm sp}_{2s}$ correctly predicts a large $S^{\rm rot}_{2s}$,
while a small $S^{\rm sp}_{2s}$ is usually obtained for small $S^{\rm rot}_{2s}$.
Besides this qualitative information,
Fig.\ref{f7} emphasizes the fact that for small
coupling strengths, the agreement between $S^{\rm sp}_{2s}$ and $S^{\rm rot}_{2s}$
is excellent, even for large admixture with other configurations,
i.e.\ for small SF.
It also shows very clearly that
the deviation becomes larger as the coupling strength $\beta$ increases.
As noted earlier, this effect is more pronounced for smaller diffuseness.

\begin{figure}
\includegraphics[width=8cm]{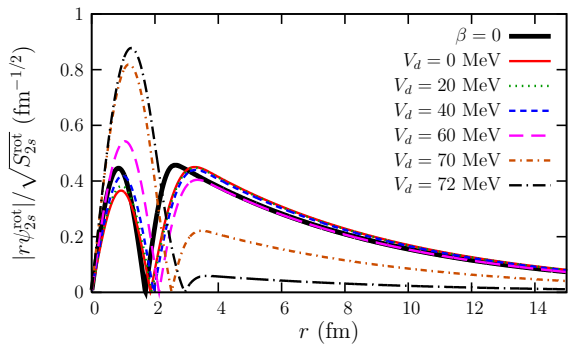}
\caption{(color online) Radial part of the $s$ overlap functions
normalized to unity obtained for $\beta=0.8$ and $a=0.1$~fm.}
\label{f7a}
\end{figure}

The largest discrepancy between the
prediction $S^{\rm sp}_{2s}$ and the ``exact'' $S^{\rm rot}_{2s}$ is
observed for small SF, where the single-particle approximation
tends to significantly underestimate the SF.
To understand this effect, we display in \fig{f7a}
the radial component of some $2s$ overlap functions obtained
in the most unfavorable case,
i.e.\ $\beta=0.8$ and $a=0.1$~fm. These functions are labeled by
the value of $V_d$ they correspond to (cf.\ \fig{f6}).
The single-particle wave function (i.e.\ $\beta=0$) is shown
as well for comparison (thick full line).
As already observed in \fig{f2}, the couplings between the configurations
affects the overlap function. In most of the cases these
changes are limited to the interior of the overlap function,
leaving the asymptotics (nearly) unchanged.
However, for some extreme cases, they extend
much beyond the range of the single-particle potential.
In particular, the node of the overlap function may be pushed so much outwards
that the ANC $C^{\rm rot}_{2s}$ nearly vanishes (see, e.g., $V_d=72$~MeV).
This leads to the very small estimates $S^{\rm sp}_{2s}$ observed in
Figs.~\ref{f4}(b), \ref{f6}(b), and \ref{f7},
whereas there remains some significant probability strength in the
interior of the wave function, i.e., a non-zero $S^{\rm rot}_{2s}$.
These rapid variations in the wave function also explain the difficulty
in adjusting the parameters of the model within the large-admixture range
(see Figs.~\ref{f4} and \ref{f6}).
However, these extreme cases seem to occur only for
a combination of large coupling strengths and significant admixture.

\subsection{Analysis with the $\delta$ coupling}\label{deltanalysis}

To expand our perspectives on the results obtained within the rotational model,
we perform a similar analysis using the delta-coupling model
developed in \Sec{delta-coupling}.
We consider here the \gs\ of a $^{11}$Be-like system.
We take into account the main component $2s$ with the core in its
$0^+$ ground state coupled to the $1d$ component with the core in
its $2^+$ excited state.
In a first step, the same square-well potential
is taken in both $s$ and $d$ partial waves.

\begin{figure}
\includegraphics[width=8cm]{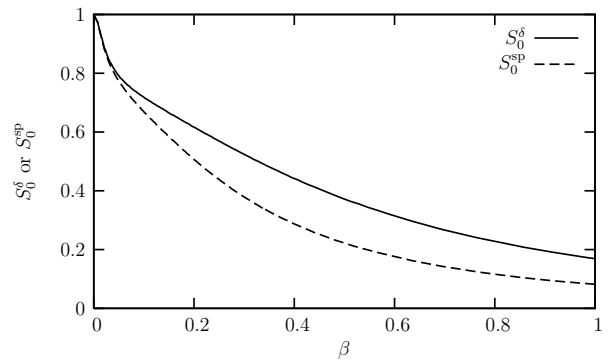}
\caption{Spectroscopic factors as a function of deformation for
the delta-coupling model.}
\label{f8}
\end{figure}

In \fig{f8} we show a plot equivalent to \fig{f1} for
the delta-coupling model. We find that $S^{\delta}_0$ (solid line)
decreases with the coupling strength $\beta$ to very small values.
Note that the quadratic dependence of $1-S^{\delta}_0$ in $\beta$
is only observed for very small coupling strength, namely $\beta<0.02$.
For $\beta>0.4$ the dominant component in
the $1/2^+$ state is no longer the $2s$ component but the $1d$ one.
This larger admixture as compared to the rotational model
may be due to the node of the $2s$.
In the rotational model, the combination of that node and
the finite extension of the coupling potential
might sufficiently reduce the source term in the $d$-wave equation
to limit the admixture. On the contrary,
the nil extension of the delta coupling avoids this cancellation effect,
which could explain the larger admixture.
Another explanation is the intrinsically larger coupling strength
in the delta coupling due to its nil diffuseness.
We have indeed seen in \Sec{Xtreme} that decreasing the diffuseness
increases the admixture between configurations
(compare Figs.~\ref{f4}(a) and \ref{f6}(a)).
Whatever the reason for that larger admixture,
it leads to a less reliable
single-particle approximation (dashed line),
as already seen in the rotational model.
The $S^{\rm sp}_0$ now
deviates significantly from the $S^{\delta}_0$ (the error can
be as large as a factor of 2).
Nevertheless, it reproduces qualitatively the general trend
of the ``exact'' SF.
Moreover, at small admixtures (i.e.\ for $S^{\delta}_0>80$\%),
the single-particle approximation is in perfect agreement with $S^{\delta}_0$.

\begin{figure}
\center
\includegraphics[width=8cm]{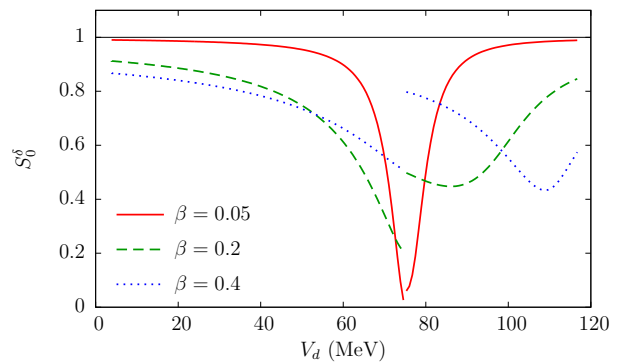}
\caption{(color online) Spectroscopic factors as a function of $V_d$ for
the delta-coupling model.}\label{f9}
\end{figure}

To confirm the degeneracy effect observed in the rotational model,
we repeat the analysis performed in \Sec{Xtreme}:
The depth of the square well in the $d$ wave $V_d$
is varied while the depth in the $s$ wave is adjusted to fit
the \ex{10}Be-$n$ binding energy.
The results of this analysis are displayed in \fig{f9},
where $S^{\delta}_0$ is plotted as a function of $V_d$
for various coupling strengths $\beta=0.05$ (full line),
0.2 (dashed line), and 0.4 (dotted line).
These results are very similar to those obtained for the
rotational model, and in particular for the small diffuseness
(see \fig{f6}(a)).
Little admixture is observed outside a range centered on $V_d\approx75$~MeV,
at which value the single-particle potential hosts a $d$ bound state
at $\varepsilon^{Bn}-\epsilon_{2^+}$.
As in the rotational model,
the width of that large-admixture range increases with the coupling strength.
We also observe discontinuities in the SF similar to those
observed in \fig{f6}(a).
These results confirm the degeneracy effect observed in \Sec{Xtreme}:
large admixture takes place when
the potential well in the coupled channel is
quantum-mechanically fit to host a bound state at the right energy.
This effect seems therefore general and not
due to some artefact of the rotational model.

\begin{figure}
\center
\includegraphics[width=8cm]{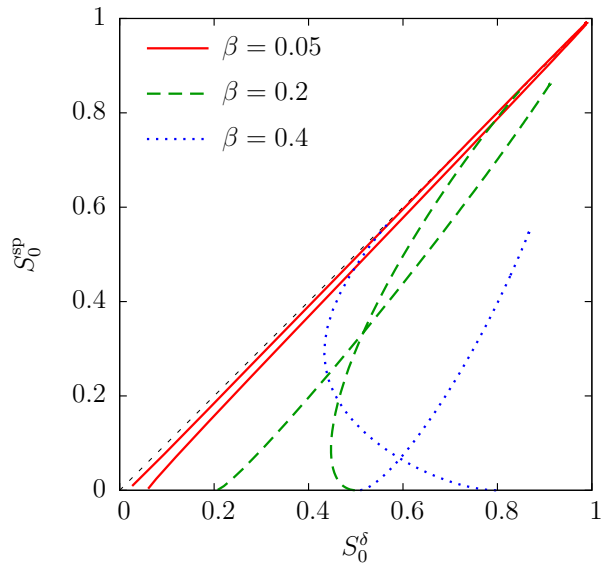}
\caption{(color online) Comparison of the single-particle approximation
$S^{\rm sp}_0$ \eq{ea12} with the SF obtained within
the delta-coupling model $S^{\delta}_0$ \eq{delta-sf}.}
\label{f10}
\end{figure}

In \fig{f10}, the single-particle approximation $S^{\rm sp}_0$ \eq{ea12}
is compared to the ``exact'' $S^{\delta}_0$ \eq{delta-sf}.
This figure is very similar to \fig{f7}. It confirms that the
single-particle approximation predicts a SF in qualitative agreement
with the ``exact'' one. The prediction is more reliable for small admixture,
i.e.\ large SF, as already seen in \fig{f7}.
At these values, the discrepancy between the ``exact'' SF
and its single-particle approximation roughly follows
\beq
S^{\delta}_0-S^{\rm sp}_0\propto 1-S^{\delta}_0,
\eeqn{e12}
in agreement with the reasoning of \Sec{delta-coupling}.
This error increases
with the coupling strength $\beta$.
Note that a similar, though less obvious, result is observed in \fig{f7}.

\section{Discussion and Conclusions}
\label{sec:summary}

Spectroscopic information about exotic nuclear structures,
like halo nuclei, can be inferred from direct reactions
\cite{Rou10,TAB07,HT03,Nak99}.
Usually a SF and/or an ANC are extracted from the analysis of
these experiments.
Most of these analyses are performed within a single-particle
approximation of the projectile wave function.
In order to evaluate the validity of this approximation,
we have studied the influence of coupling between configurations
upon the overlap wave function.
Keeping in mind the extraction of SF from peripheral reactions
\cite{MN05,CN07}, we have focused our study on the effects of
these couplings on the SF and the ANC.
Assuming that an ANC can be reliably extracted from direct reactions,
we have analyzed how accurately a SF can be deduced from it.

In this work, we have considered
a collective model of the nucleus in which a valence neutron is
bound to a deformed core allowed to excite.
In particular, we have used the
rotational-coupling model of Nunes \etal \cite{NTJ96,face},
in which the core is described as a deformed rotor.
In this model the various states of the core are assumed to be
part of a rotational band.
A qualitative delta-coupling model has also been used to investigate
the results of the rotational model.
Calculations have been performed in the case of the typical one-neutron halo
nucleus \ex{11}Be.

Interestingly, this analysis shows that
even a small coupling may affect significantly the overlap function.
However, the changes compared to the single-particle wave function
appear mostly in the interior,
leaving the asymptotics nearly unchanged (see \fig{f2}).
This surprising result suggests that probing only the tail
of the wave function may give a reasonable estimate of the SF, contrarily
to what has been assumed in \Ref{CN07}.
To explore this idea,
we vary the deformation of the core that acts as a coupling strength.
The single-particle approximation \eq{ea12} is then compared
to the ``exact'' SF obtained within the rotational model.
This approximation is obtained from
the ratio between the ANC of the rotational-coupling model
and that of the single-particle model. 
The former acts as the actual ANC of the system,
supposedly measured from peripheral reactions.
The latter corresponds to the one obtained with a single-particle
description of the nucleus, which is used in most of the reaction models.
The prescription \eq{ea12} therefore simulates the procedure performed
in extracting SFs from experimental data.

The results of our analysis show that the single-particle approximation
often provides a reliable estimate of the SF.
This estimate is at least in qualitative agreement
with the coupled-channel models: It predicts large
(respectively small) SFs, when large
(respectively small) SFs are obtained from the coupled-channel models.
For very small coupling strength, this estimate is very accurate.
On the contrary, for large coupling strength, and in particular
for large admixture of the configurations, i.e.\ small SF,
the single-particle approximation is much less reliable.

Being obtained within two independent structure models,
this result seems quite general.
To understand it on firmer grounds, let us look back at the
coupled equations satisfied by the overlap function \eq{ea8}
and compare them with the single-particle equation \eq{ea10}.
The overlap function $\psi_{\nu_A\nu_B}$ will depart
from its single-particle approximation only when the coupling terms
in \Eq{ea8} become significant, i.e.\
when at least one of the products $V_{\nu_B\mu_B}\psi_{\nu_A\mu_B}$ is large.
This will happen for large coupling strengths, and/or
when the population of the $\nu_A\mu_B$ configuration is large
in the range of $V_{\nu_B\mu_B}$,
i.e.\ when there is a significant admixture between the various
channels. In these two cases, the coupling affects
the overlap function so much that the single-particle approximation
no longer holds.
This reasoning also illustrates the possible influence of a
node in one of the overlap function upon the admixture.
Such a node may indeed cancel out the effect of the coupling
if it is located within the range of $V_{\nu_B\mu_B}$.
In the future, we plan to further investigate this effect.

This study suggests the simple rule of thumb that
large SFs deduced from ANCs are reliable.
On the contrary, single-particle analysis of data
estimating small SF, i.e. a large admixture between configurations,
should not be relied on.
Note that in any case some uncertainty remains, and that even large
SF estimates can be off by 10--20\%.
Therefore, these estimates must be taken for what they are---estimates---and not
precise, unquestionable values.

The accuracy of the
single-particle approximation depends also on the geometry of the
potential chosen to simulate the mean field of the core.
In practice this geometry is unknown.
We have not considered this uncertainty in the present analysis.
Fortunately, as observed by Sparenberg \etal \cite{SCB10},
the single-particle ANC for loosely-bound states
is not strongly dependent on the potential geometry.
In particular for an $s$ valence neutron, it can be efficiently estimated
from only the binding energy of the system (see Eq.~(20) of \Ref{SCB10}).

In our study, we consider a collective model to simulate microscopic effects.
To evaluate the sensitivity of our conclusions to this approximation,
it would be interesting to repeat this analysis with microscopic models,
such as the microscopic cluster model \cite{BDT94},
the fermionic molecular dynamics \cite{NFR05,NF08},
or the Green's function Monte Carlo model \cite{PW01},
which all can correctly
describe the asympotics of overlap wave functions for loosely-bound
nuclear systems.

\begin{acknowledgments}
P.~C. acknowledges the support of the Fund for Scientific Research
(F.~R.~S.-FNRS), Belgium.
This work was also partially supported
by the National Science Foundation grant PHY-0800026 and
Department of Energy grant DE-FG52-08NA28552.
This text presents research results of the Belgian Research Initiative
on eXotic nuclei (BriX), program P6/23 on interuniversity attraction
poles of the Belgian Federal Science Policy Office.

\end{acknowledgments}

\appendix

\section{Effective Hamiltonian in a single-particle equation}
\label{appendix_Ham}

Given the linearity of the set of equations \eqref{ea8} for the overlap functions, any overlap function is linear in any other of those functions.  Any selected equation may be then formally closed by expressing other overlap functions in terms of the function of interest.  The cost is in a potentially strong nonlocality, both in position and energy, for the effective Hamiltonian within the single-particle equation for the chosen overlap function.  Also, it may be difficult to construct the effective Hamiltonian {\em in practice}.

Let the chosen overlap function correspond to the state $\nu_B$ of the core $B$.  First,
energy eigenvectors need to be found in the remaining space of overlap functions, in the absence of coupling to $\nu_B$.  With different energy values and eigenvectors denoted with index $\alpha$, an eigenvector $\eta^{\overline{\nu}_B \, \alpha}$ solves the set for energy $E^{\overline{\nu}_B \, \alpha}$,
\begin{align}
\big( E^{\overline{\nu}_B \, \alpha} - \epsilon_{\mu_B} \big) \, \eta_{\mu_B}^{\overline{\nu}_B \, \alpha} = t_{\ve{r}}
\, \eta_{\mu_B}^{\overline{\nu}_B \, \alpha} & +   \sum_{\gamma_B \ne \nu_B} V_{\mu_B \, \gamma_B} \, \eta_{\gamma_B}^{\overline{\nu}_B \, \alpha} \, , \notag \\ & (\text{for} \; \mu_B \ne \nu_B) \, .
\label{eq:nonu}
\end{align}
Note that, without the coupling potentials $V_{\mu_B \, \gamma_B}$, the states $\lbrace \eta^{\overline{\nu}_B \, \alpha} \rbrace$ would just represent energy eigenstates within the single-particle potential, coming in near-degenerate sets, because of the expected weak-dependence of the potential on $\mu_B$.
Given the coupling to $\psi_{\nu_A\nu_B}$ of the overlap functions $\psi_{\nu_A \, \mu_B}$, where $\mu_B \ne \nu_B$, the set of the latter functions can be next expressed as a combination of the vectors~$\lbrace \eta^{\overline{\nu}_B \, \alpha} \rbrace$.  Upon inserting the expressions for the functions $\psi_{\nu_A \, \mu_B}$ into the equation for the overlap function $\psi_{\nu_A\nu_B}$, we arrive at the closed equation:
\begin{align}
\big(\varepsilon_{\nu_A}^{Bn} -  \epsilon_{\nu_B} \big) & \, \psi_{\nu_A\nu_B}({\pmb r})  =  \Big(  t_{\ve{r}}  + V_{\nu_B \, \nu_B}({\pmb r})   \Big) \, \psi_{\nu_A\nu_B}({\pmb r}) \notag \\ &
 + \int d{\pmb r}' \, \Delta V_{\nu_B \, \nu_B} ({\pmb r}, {\pmb r}'; \varepsilon_{\nu_A}^{Bn}) \, \psi_{\nu_A\nu_B} ({\pmb r}') \, ,
\label{eq:muonly}
\end{align}
where the modification of the single-particle potential is
\begin{align}
\Delta V_{\nu_B \, \nu_B} ({\pmb r}, {\pmb r}'; E) = & \sum_{\substack{\mu_B \ne \nu_B \\ \gamma_B \ne \nu_B}}
V_{\nu_B \, \mu_B} ({\pmb r}) \notag \\ & \times  G_{\mu_B \, \gamma_B}^{\overline{\nu}_B} ({\pmb r}, {\pmb r}'; E) \, V_{\gamma_B \, \nu_B} ({\pmb r}') \, ,
\label{eq:DV}
\end{align}
and the Green's propagator within the space complementary to $\nu_B$ is
\begin{align}
G_{\mu_B \, \gamma_B}^{\overline{\nu}_B} ({\pmb r}, {\pmb r}'; E) = & \sum_\alpha \frac{1}{E - E^{\overline{\nu_B} \alpha}} \notag \\
& \quad \quad \times
 \eta_{\mu_B}^{\overline{\nu}_B \, \alpha}({\pmb r}) \, \left[ \eta_{\gamma_B}^{\overline{\nu}_B \, \alpha } ({\pmb r}')
 \right]^* \, .
\label{eq:GE}
\end{align}
The results are written here assuming local 2-body interactions.

The nature of the contributions to the Green's function, and to the modification of the potential $\Delta V$, from different excitations of the core, will depend on the core excitation energy.  High-lying excitations will yield contributions where propagation both to the outside and inside of the nucleus represents a tunneling.  Those contributions will not yield much nonlocality for $\Delta V$, neither as a function of position nor energy.  On the other hand, for low-lying excitations, the propagation within the nuclear volume may be largely uninhibited, while still turning to tunneling in the exterior.  The latter excitations will contribute nonlocalities to $\Delta V$, both in terms of energy-dependence and position, with latter nonlocalities possibly extending across the nucleus.  The different limits of nonlocality may be, in particular, easily assessed in the delta-coupling model.  In this paper, the focus is on the impact of coupling to low-lying states, on the overlap functions and on the deduced SFs.  We may mention, though, that many microscopic interactions $v_{ij}$ with strong short-range repulsion may yield unphysical single-particle potential $V_{\nu_B \, \nu_B}$ in absence of any modification in the form of $\Delta V$.  For such interactions, some level of renormalization of the potentials $V_{\nu_B \, \mu_B}$, for low-lying states, through coupling to high-lying states may be necessary to start with.  The consequence in the possible ill-definition of SFs, due to the latter type of couplings, has been indicated in Ref.~\cite{FH02}.


\end{document}